# A POSSIBLE ASTRONOMICALLY ALIGNED MONOLITH AT GARDOM'S EDGE

## DANIEL BROWN, ANDY ALDER, ELIZABETH BEMAND

A unique triangular shaped monolith located within the Peak District National Park at Gardom's Edge could be intentionally astronomically aligned. It is set within a landscape rich in late Neolithic and Bronze Age remains. We show that the stone is most likely in its original orientation owing to its clear signs of erosion and associated to the time period of the late Neolithic. It is tilted towards South and its North side slopes at an angle equal to the maximum altitude of the Sun at mid-summer. This alignment emphasizes the changing declinations of the Sun during the seasons as well as giving an indication of mid-summers day. This functionality is achieved by an impressive display of light and shadow on the North-facing side of the Monolith. Together with other monuments in the close vicinity the monolith would have represented an ideal marker or social arena for seasonal gatherings for the else dispersed small communities.

*solar alignment, monolith, late Neolithic, seasons, shadow casting*

## Introduction

This paper is concerned with the possibility of a standing stone located in the Peak District National Park being astronomically aligned. Single standing stones (monoliths) are uncommon in this area; some may have been lost for gate posts or building materials. Overall they appear more widespread in the western counties of the United Kingdom (Burl 1995, 49).

There are only a few sites of singular standing stones in the Peak District including: Wirksworth Monolith, Manystones Lane, and Gardom's Edge. The monolith at Gardom's Edge on the Eastern Moors Estate is sited in an area of much early human activity ranging from a low density of Mesolithic microliths, rock art, Neolithic enclosures, and Bronze Age field systems. A more detailed overview of the Gardom's Edge archaeological landscape is given in Barnatt et al. (2002). It is likely that the stone is contemporary with the rock art or erected during the establishment of early Bronze Age ritual monuments that have been found across the Eastern Moors Estate associated with periods of cultivation. This allows for an approximate date of the monolith of 2,500 – 1,500 BC (Barnett and Smith, 2004, 50). The 2.2m high Gardom's stone consists of sandstone with a lithology known locally as millstone grit, medium to coarse grained with moderately well sorted sub angular grains. In its current position sedimentary planes within the stone are near vertical, the east and west face of the monolith being bedding planes. These planes are perpendicular to sedimentary structures in natural outcrops in the area. The main stone has a smaller sister stone identified in Barnatt et al. (2002), approximately 90 m North-West, though this stone is fairly nondescript. The interesting factor of the Gardom's stone that the stone appears to have been selected for its triangular shape and that it appears to be packed at the base on one side (Figure 1), suggesting man made alignment.

## The stone within its context

A standing stone such as this may predate the surrounding settlements and be an example of a focal point that small communities would gather at seasonally, as Edmonds and Seaborne (2005) state 'Work has also suggested that for some stones at least, there was a seasonal pattern to the encounter that people would have had.'

Gardom's Edge consists of a high plateau of typical altitude for this region of the Peak District overlooking the Derwent valley below and contains a distinctive gritstone scrap clearly visible from below. These factors could have made this an easy location to find. Although pollen evidence suggests that forests dominated the

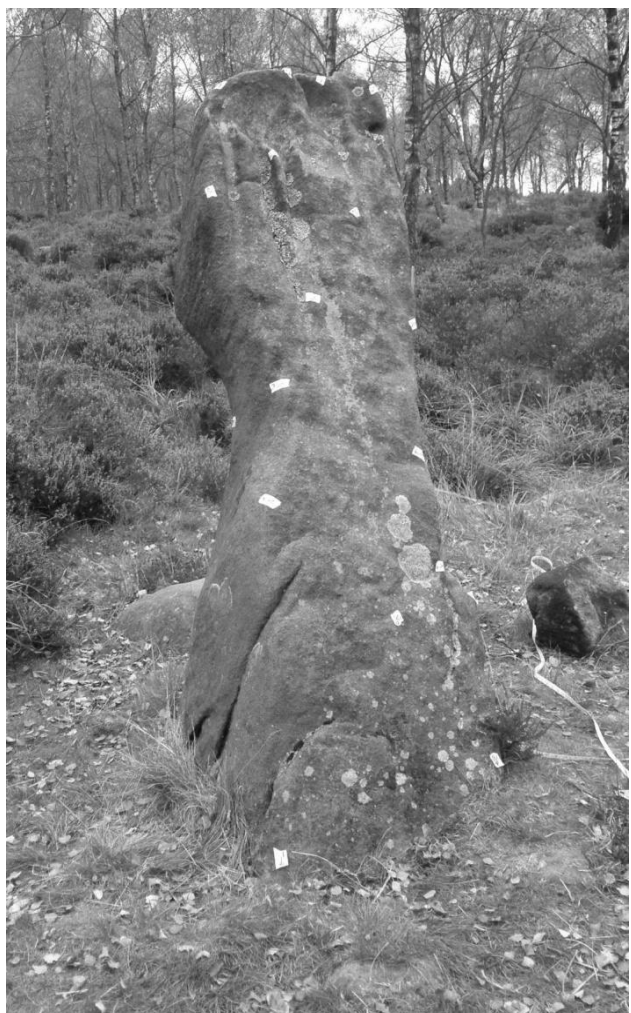

*Figure 1. Image taken of the standing stone at Gardom's Edge during three-dimensional survey showing the flattest north facing side. The visible white markers are reference points applied to the surface. This side shows clear signs of erosion. At the base of the stone several packing stones can be seen.*

Eastern Moors (Edmonds and Seaborne, 2010) in the Neolithic, Pryor (2010) points out that we should not believe that during these times all England was densely wooded; but high moorlands e.g. in Cornwall would have been treeless. Gardom's edge is such a high moorland and patches of it could have been free of ancient woodlands allowing a clearer view of the standing stone at Gardom's.

If the stone is contemporary to the rock art and the land was also settled at the time it could be argued that that it was either astronomically aligned or a marker seen on a daily basis (Edmonds and Seaborne, 2005). However the position and shape of the stone points to a more designed use giving the first scenario a logical credence.

**Alignment survey and the seasonal sundial model**
The standing stone presents a rather unusual triangular shape that seems to highlight a direction. The north facing side shown in Figure 1 is rather flat with any structure only caused by erosion from rainwater. A detailed survey was carried out at the end of July 2010. Figure 2 gives the location at which gradient and strike were measured. The strike is given as degrees from geographic North. The detailed survey revealed the alignment of the stone towards South $(92.0\pm2.1)°$, but also an overall gradient of the north face of $(58.3\pm2.9)°$. The maximum altitude of the Sun ($A_{max}$) for the approximate time of erection is $60.7°$. This altitude has been derived given the location of the stone at a geographic latitude of $\varphi=53.26°$ North and the obliquity of the ecliptic $\varepsilon=23.95°$ calculated from Wittmann (1979). Figure 3 illustrates how $A_{max}$ was derived using the obliquity of the celestial equator plane (apparent solar path during the equinox) and the apparent solar path during the summer solstice. The plane defined by the north facing side of the stone is indicated by the tilted transparent grey plane. Both gradient and $A_{max}$ are comparable and have led us to believe that the standing stone at Gardom's Edge could have been astronomically aligned marking the passage of seasons during the year.

Such an astronomically aligned stone could be described as a seasonal sundial described in the following. However it is not intending to mark local time during a day or measure exact dates during a year. Rather the seasonal shadow casting allows for the display of cosmological knowledge such as the 'death' and 'rebirth' of the Sun leading to dramatic ceremonies. The principle of a seasonal sundial is modelled in Figure 4 where three panels illustrate the apparent passage of the Sun over the site of the standing stone during the winter solstice, summer half of the year, and summer solstice. The north facing side will be illuminated if the Sun is above this plane. If the path falls below the plane of the north facing side of the stone the line is solid and if it falls above the plane it is dashed. As can be seen in II of Figure 4 the Sun can cross the plane up to two times, marked here as points B and C. These points move closer together as the summer solstice approaches. If the plane has an obliquity regarding the horizon of less than $90-\varphi+\varepsilon$ points B and C will never merge and the north facing side will never be illuminated during midday. If it is exactly $90-\varphi+\varepsilon$ the points will merge to one and the north facing side will always be illuminated on the day of the Summer solstice as illustrated in III of Figure 5. A lower tilt of the north facing side will extend the illumination period to several days before and after the summer solstice. To achieve such behaviour of the shadow, the flat side of the stone has to be orientated towards North, i.e. the plane intersects the horizon in an E-W direction.

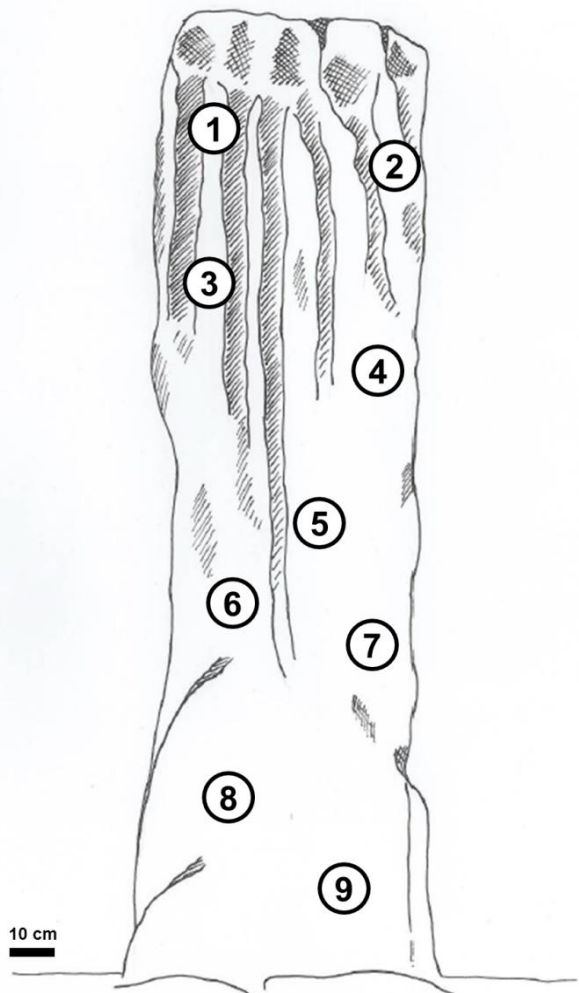

*Figure 2. Gradient survey of the north facing side of the standing stone. Nine locations were surveyed with regards to their gradient and strike perpendicular to the gradient. Each location is marked and gives an average gradient $(58.3\pm2.9)°$ of and a strike of $(92.0\pm2.1)°$.*

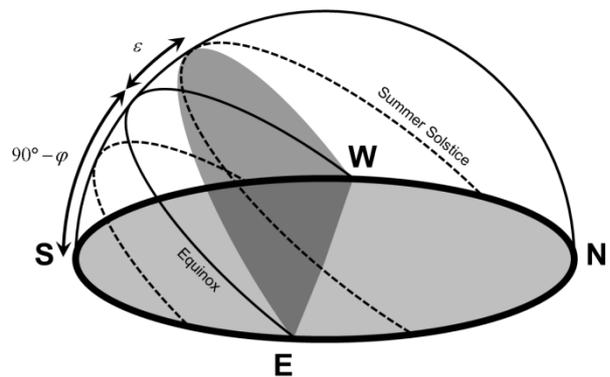

*Figure 3. This sketch illustrates three apparent paths of the Sun during a day either at the Solstices (dashed) or at the Equinoxes (solid). The equinox path is equivalent to the celestial equator. Assuming a geographic latitude of $\varphi$ the path has an obliquity with regards to the local horizon of $90 - \varphi$. The maximum solar altitude ($A_{max}$) achieved during the summer solstice is determined by increasing the obliquity by $\varepsilon$ to a total of $90 - \varphi + \varepsilon$. The plane defined by the north facing side of the stone is shaded in grey.*

Given this behaviour, a seasonal sundial indicates the winter half of the year by having its north facing side cast in permanent shadow.

Only after the equinoxes will the north side become partly illuminated during the mornings and evenings. When the time of the summer solstice approaches the north facing side will be illuminated during the entire day. The number of hours for which it still remains in shadow is given in Figure 5 and have been derived with the appropriate obliquity of the ecliptic. The hashed area indicates permanent illumination and the solid black area permanent shadow. The number of weeks the north facing side will be illuminated close to the solstice is given in Figure 6.

To further model the shadow on the standing stone at Gardom's Edge using its real shape, an additional three dimensional survey was carried out in October 2010 (see Figure 1). A rendered model of the standing stone is shown in IV in Figure 5 and shows the shadow at local midday on May 10. The north facing side has just been illuminated for the first time which is earlier than predicted in Figure 5 but this was expected given the more realistic relief of the North facing side.

**Examples of Prehistoric shadow casting**

The astronomical knowledge of seasonal shadow casting was already well developed in prehistory (see Ruggles 1999, 130). Also the practical and simple set-up of the seasonal sundial model proposed for the interpretation of the astronomical alignment supports the more hands-on approach to astronomy in prehistory, rather than a more abstract visualisation of geometry and time (Ruggles 1999, 79) as used in more familiar horizontal sundial with a gnomon.

Within the context of the British Isles there are two possible other examples of intentional use of shadows that are associated to a period around 2000 BC. Both are not local to Gardom's Edge and are incorporated into burial monuments.

Pendergast (1991) reports that carving patterns found on the kerbstone (K1) at Newgrange follow seasonal shadow movements caused by standing stones in the great circle (GC1 and GC2). These could indicate a possible calendar intention included into Newgrange.

Additionally, Trevarthen (2001) found indications that the standing stones surrounding some Clava cairns

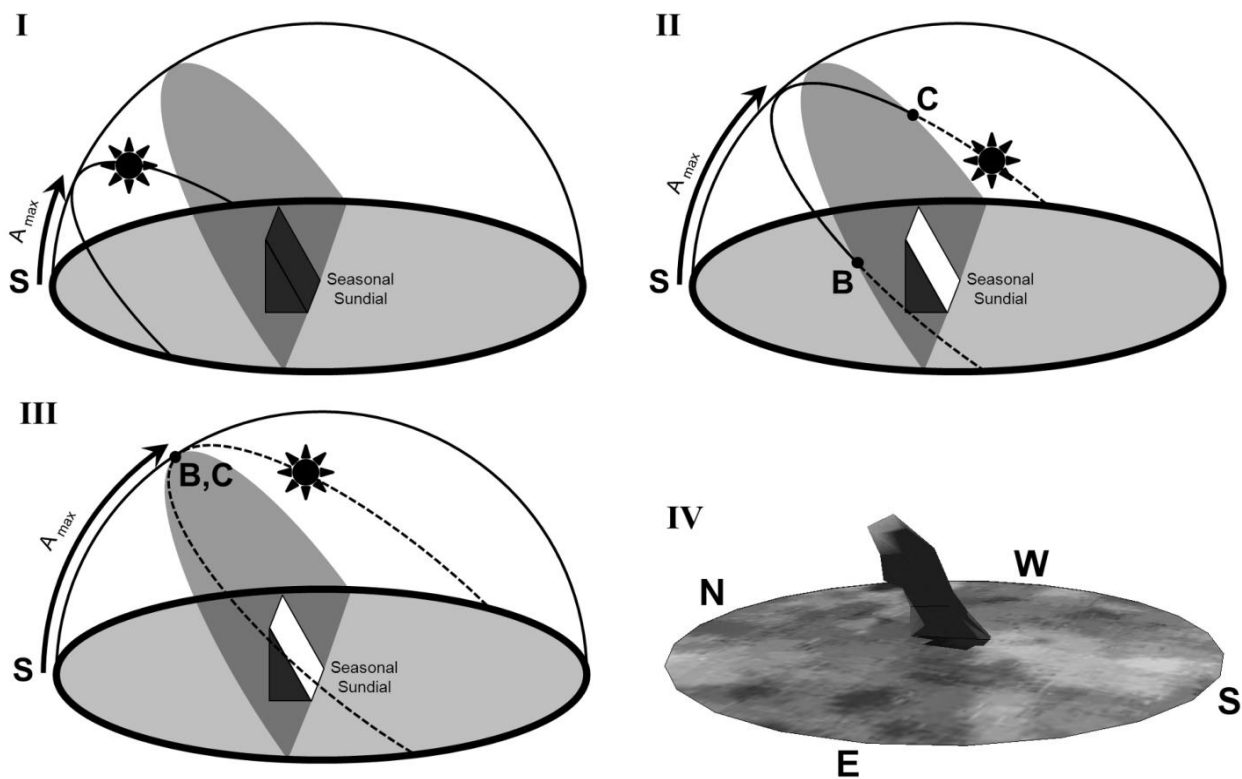

*Figure 4. The principle of a seasonal sundial is illustrated in panels I, II, and III. The apparent path of the Sun during the winter solstice (A), an arbitrary day in the summer half of the year (B), and the summer solstice (C) is shown by the black arc with the symbol of the Sun. If the Sun is below the plane shaded in grey defined by the north facing side of the seasonal sundial, the arc is solid. If the Sun is above the arc is dashed. The location at which the arc of the Sun intersect the plane is indicated by points B and C. In panel IV a three dimensional render of the standing stone is shown for local midday on May 10. The model shows the first illumination of the north side during midday for a more realistic shadow on the stone .*

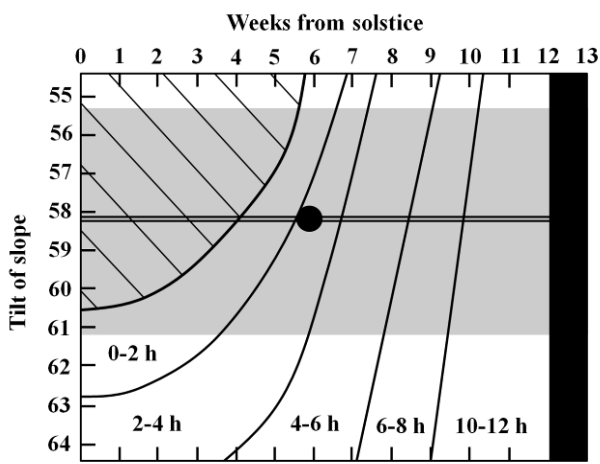

*Figure 5. This contour map of the tilt of the plane in degrees against weeks from solstice illustrates the amount of hours the north facing side of the stone remains in darkness around midday. The shadow interval is binned into two hour segments, the hashed area shows times when the north side is permanently illuminated and the black area illustrates times of permanent shadow. A geographic latitude and epoch for the obliquity are chosen according to the site and erection date. The surveyed dip of the north facing side is marked by a horizontal double line and the error range is indicated by the shaded area. The first illumination of the north side in the rendered model is marked by a black point.*

(Balnvaran Clava) in the north-east of Scotland could have cast shadows onto the central cairn to mark a seasonal calendar. It is noted that this incorporation of cyclic time into a burial monument is an important symbol representing eternity.

**Discussion**
The intentional astronomical alignment of the Gardom's Edge standing stone during 2500 – 1500 BC can only be confirmed through its alignment and dip of the north facing side alone. Given its uniqueness as one of the few single standing stones in this region, this idea cannot be confirmed through comparison with other sites close by. This fact makes it challenging to rule out chance alignment of the stone that could seem to create a seasonal sundial.

However, the presence of the packing stones indicates intentional erection. Furthermore, the stone is highly unlikely to be a glacial eratic, since it is typical to the local bedrock lithology and has a highly angular shape. The unusual shape of the stone strengthens the choice of stone towards its specific purpose.

Other examples of shadow casting in the British Isles have also demonstrated that the skills were present at this time including the symbolic importance. Even though at Gardom's Edge this is not enforcing a burial monument, it is included into a site and possible ritual environment that is encountered and used seasonally.

The stone itself seems not to have been moved, since the packing stones are still in place and it fits into the general context of the ancient history of the overall site. But visual inspection and a weathering survey have revealed effects of possible subsidence and erosion.

| Date | δ degrees | Tilt degrees |
|---|---|---|
| Solstice | 24.0 | 60.7 |
| 1 week | 23.8 | 60.5 |
| 2 weeks | 23.2 | 60.0 |
| 3 weeks | 22.3 | 59.1 |
| 4 weeks | 21.1 | 57.8 |
| 5 weeks | 19.6 | 56.3 |
| 6 weeks | 17.7 | 54.5 |
| 7 weeks | 15.7 | 52.4 |
| 8 weeks | 13.4 | 50.1 |

*Figure 6. This table lists the astronomical declination, δ, of the Sun in degrees for times from the summer solstice. It also lists the according tilt in degrees of the north facing side of a seasonal sundial model to achieve permanent illumination in the given period of time around the summer solstice.*

The stone seems to lean towards West as can be seen in Figure 2 which has been estimated to about (4±4)°.
If the stone was set up with a tilt, or if even the north facing side of the stone had a slightly steeper tilt cannot be determined at this time. Further studies on the structure of the packing stones and the general strata at the base of the stone are made very difficult since they may cause subsidence to the stone. Future non-invasive studies are proposed. An increased dip of the north facing side by 1–2° would not rule out its usefulness but actually improve the astronomical alignment towards the theorised seasonal sundial, reducing the interval of full illumination as shown in Figure 5.

Erosion has clearly shaped this stone and we found karstic surface features such as solution pits, solution runnels and decantation flutings, produced by rainwater pooling on flat surfaces and flowing down rock faces. The type and intensity of these features is influenced by the orientation of the rock face while these erosional processes are in action.

The systematic mapping of the type, scale, and intensity of weathering forms provides an indication of the time scale over which dissolution processes have acted in their current orientation.

Solution pits have developed across the flat surface at the top of the sandstone, rounded and sub rounded in shape with sub vertical to convex sidewalls and irregular floors. Pits are laterally linked with sandstone interfluves forming saddle shaped ridges. Decantation flutings dissect the sub vertical surfaces forming vertical channels on the south face of the stone. Solution runnels dissect the inclined surface of the north face of the stone (see Figure 1 and 2). The runnels run in parallel down the slope and have a U-shaped channel like cross-section.

The systematic mapping of weathering features indicates that the stone has been at its current orientation for a significant period. All dissolution features are consistent with erosional processes at this orientation, perpendicular to natural outcrops in the area and have also occurred in other Neolithic stone circles consisting of sandstone (see Duddo stones NT931437; Younger and Stunell, 1995).

**Conclusion**

The finding presented here allows for the possibility that the standing stone at Gardom's Edge was astronomically aligned during the late Neolithic and early Bronze Age period. Therefore, this standing stone would have represented an ideal marker or social arena for seasonal gatherings for the else dispersed small communities since it incorporates seasonal shadow casting within its design. This fact ties in with the idea that landscapes can provide the "lore of life" including standing stones presented by Pryor (2010).

We have found that the shape and orientation of the stone are unique and point towards an intentional erection. Furthermore, we find that the weathering evidence and packing stones suggest that the stone has not been moved; therefore allowing us to assume orientation and alignment to be genuine. The present subsidence will only marginally have altered the gradient of the north facing side of the stone. Erosion has been substantial but localised into runnels that were avoided during the survey (see Figure 2), allowing its effects to be neglected.

However, since this is a unique astronomical use of such a rare standing stone, possible chance alignments cannot be ruled out. Further in-depth study of this standing stone would be required to confirm its possible astronomical alignment.

**Acknowledgements:**

DB would like to thank the pre-16 work placement students W. Busby, N. Sikand-Youngs and R. Loyd Miller contributing to the three dimensional model of the standing stone. Furthermore, DB is extremely grateful to the feedback provided by B. Bevan and the discussion with F. Pendergast at SEAC 2012 that have improved the presented paper.